\documentclass[usenatbib]{mn2e}

\usepackage{psfig,url}

\footnotesize
\normalsize

\title[Eight New LMC Pulsars]
{Eight New Radio Pulsars in the Large Magellanic Cloud}
\author[Ridley et al.]
{J.P.~Ridley$^{1}$, F.~Crawford$^{2}$, D.R.~Lorimer$^{3,4,5}$, S.R.~Bailey$^{6}$, J.H.~Madden$^{2}$, 
\newauthor R.~Anella$^{2}$ and J.~Chennamangalam$^{3}$\\
\\
$^1$Department of Engineering and Physics, Murray State University, 
Murray, KY~42071, USA\\
$^2$Department of Physics and Astronomy, Franklin and Marshall College, 
Lancaster, PA~17604, USA\\
$^3$Department of Physics, West Virginia University, 
PO~Box~6315, Morgantown, WV~26506, USA\\
$^4$National Radio Astronomy Observatory, PO Box 2, 
Green Bank, WV~24944, USA\\
$^5$Astrophysics, University of Oxford, Denys Wilkinson Building, Keble Road, Oxford OX1 3RH\\
$^6$Aerospace Engineering Program, Syracuse University, 263 Link Hall, Syracuse, NY 13244, USA\\
}
\date{Accepted 2013 April 23.  Received 2013 April 23; in original form 2013 March 7}
\begin{document}
\maketitle
\newcommand{\setthebls}{
}
\setthebls

\begin{abstract} 
We present the discovery of eight new radio pulsars located in the
Large Magellanic Cloud (LMC). Five of these pulsars were found from 
reprocessing the Parkes Multibeam Survey of the Magellanic
Clouds, while the remaining three were from an ongoing new survey 
at Parkes with a high resolution data acquisition system.  It is possible 
that these pulsars were missed in the earlier processing due to radio 
frequency interference, visual judgment, or the large number of candidates 
that must be analysed.  One of these new pulsars has a dispersion measure 
of 273~pc~cm$^{-3}$, almost twice the highest previously 
known value, making it possibly the most distant LMC pulsar.  In addition, 
we present the null result of a radio pulse search of an X-ray point source 
located in SNR J0047.2$-$7308 in the Small Magellanic Cloud (SMC).  Although no 
millisecond pulsars have been found, these discoveries have increased the 
known rotation powered pulsar population in the LMC by 
more than $50\%$. Using the current sample of LMC pulsars, we used a 
Bayesian analysis to constrain the number of potentially observable pulsars 
in the LMC to within a 95\% credible interval of 
57000$^{+70000}_{-30000}$.  The new survey at Parkes is $\sim$20\% complete 
and it is expected to yield at most six millisecond pulsars in the LMC and SMC.  
Although it is very sensitive to short period pulsars, 
this new survey provides only a marginal increase in sensitivity to long periods.  The 
limiting luminosity for this survey is 125~mJy~kpc$^2$ for the LMC
which covers the upper 10\% of all known radio pulsars.
The luminosity function for normal pulsars in the LMC is consistent with
their counterparts in the Galactic disk. The maximum 1400~MHz radio
luminosity for LMC pulsars is $\sim 1000$~mJy~kpc$^2$. 
\end{abstract}


\section{INTRODUCTION}\label{sec:intro}

Our nearest galactic neighbours, the Large and Small Magellanic
Clouds (LMC and SMC), are irregular galaxies with a different star
formation history than the Milky Way. They therefore represent very
interesting targets for pulsar searches.  Previous pulsar surveys of
the Magellanic Clouds by \citet{mha+83}, \citet{mmh+91},
\citet{ckm+01}, and \citet{mfl+06} have discovered 13 radio pulsars in
the LMC.  In addition, two rotation-powered X-ray pulsars have been discovered (see
\citet{ssh84} and \citet{mgz+98}) leading to a total of fifteen known
rotation-powered LMC pulsars.  Five radio pulsars were also discovered in
the SMC in the surveys by \citet{mmh+91}, \citet{ckm+01}, and
\citet{mfl+06}, yielding a total of twenty rotation powered pulsars in the
Magellanic Clouds.

Pulsar surveys with the Parkes 20-cm multibeam receiver \citep{swb+96}
have been carried out along the plane of the Galaxy \citep{mlc+01}, at
intermediate \citep{ebsb01} and high Galactic latitudes
\citep{jbv+03,bjd+06}, and in the Magellanic Clouds \citep{mfl+06}.  These 
surveys have been very successful.  Improved computer processing 
capabilities and search algorithms have inspired a reprocessing of these 
various Parkes Multibeam Surveys, most notably the Galactic plane survey, 
and have led to a large number of additional discoveries
\citep{kel+09,mlb+12,eklk13,kek+13}. In this paper, we present the
discovery of five new radio pulsars in the LMC from a reanalysis of the
\citet{mfl+06} survey data and three from a high resolution survey of the 
LMC which we are conducting at Parkes for a total of eight new radio pulsars.

In Section~\ref{sec:searches}, we introduce the searches that were
performed.  In Section~\ref{sec:reduction}, we explain the data
reduction stage, including the dispersion measure (DM), period ($P$),
and acceleration search ranges.  Our results are then presented in
Section~\ref{sec:results}.  We discuss the implications of our results
in Section~\ref{sec:discussion}, and finally, in
Section~\ref{sec:conclusions}, we draw our conclusions.

\section{OVERVIEW OF THE SEARCHES}\label{sec:searches} 

Five separate searches using both new and archival data were performed
in order to detect new radio pulsars in the SMC and LMC.  Data from
the Parkes Multibeam Survey of the Magellanic Clouds \citep{mfl+06} were 
reprocessed and searched using three different sets of search parameters.  
The first pass focused on low dispersion measures (up to 300~pc~cm$^{-3}$) 
and used a broad, coarse acceleration search.  The second trial extended 
the DM range to 800~pc~cm$^{-3}$ and had a very fine acceleration search.  
The third search used the archival data to target beams that contained 
high-mass X-ray binaries (HMXBs) in the LMC.  These survey beams were then 
reprocessed using a very intensive acceleration search.  New data were 
processed from a deep search of an X-ray point source in the SMC and a new 
high resolution survey of the LMC using the Parkes Telescope.

\subsection{Parkes Multibeam Survey}

Archival data from the Parkes Multibeam Survey of the LMC and SMC 
were reprocessed and searched by each team.  The original 
observations took place between May 2000 and November 2001.  
Using a centre frequency of 1374~MHz and a bandwidth of 288~MHz split 
into 96 channels, data were sampled every 1~ms for a total observing 
time of 8400~s per pointing \citep{mfl+06}.  Due to increased computing 
power and search algorithms, allowing us to use lower signal-to-noise 
ratios and higher acceleration ranges (see Section~\ref{sec:reduction}), 
we were sensitive to lower luminosity and more highly accelerated 
pulsars than the \citet{mfl+06} survey.

\subsection{High-mass X-ray Binaries}

The Magellanic Clouds have 128 catalogued HMXBs \citep{lph05}, and this 
provided a large sample of sources to
search for corresponding radio pulsar signals with a uniform
sensitivity.  Multiwavelength detection of a HMXB as a radio pulsar 
provides insight into the evolution of these systems and into the 
physics of the companion star (see, e.g., \citet{tbep94} and 
\citet{gtp+95} pertaining to PSR B1259$-$63).

Despite extensive study of these HMXB systems (both in the Galaxy and
in the Magellanic Clouds) at X-ray wavelengths as both persistent and
as transient/bursting sources, only four radio pulsars have been
found to date to be orbiting similarly high-mass, non-degenerate
companions: PSRs J0045$-$7319 (itself in the SMC, \citet{kjb+94}), J1638-4725 \citep{l08}, 
J1740$-$3052 \citep{sml+01}, and B1259$-$63 \citep{jml+92}. Of these, only 
PSR B1259$-$63 is a member of a Be/X-ray binary system.  Is it the case 
that accretion and/or eclipsing from the donor prevents radio emission 
from being detectable in many cases?  Perhaps, but some HMXBs may be 
detectable as radio pulsars during the non-accretion phases of their 
orbits, if their orbits are sufficiently wide.  PSR~B1259$-$63 is one of these
wide-orbit, eccentric systems, and it has been extensively studied as
both a radio pulsar and transient X-ray source since its discovery
more than twenty years ago \citep{jml+92}, particularly at or near its
periastron passages, which occur every 3.4 years. This system
continues to provide exciting results using multiwavelength
observations (see, e.g., \citet{cna+09}, \citet{kabr12}, and \citet{kch12}).

HMXBs from the Magellanic Clouds High-Mass X-Ray Binaries 
Catalog (HEASARC) \citep{lph05} were identified and checked for coincidence 
with the existing beams in the MC radio survey.  These beams were then 
selected to run a directed search for radio emission using the Parkes 
Multibeam Survey data.  Of the 128 HMXBs in the catalog, 36 had positions 
that were either not in a coincident beam or found near the edge of a beam 
(therefore yielding low sensitivity), 16 were found in beams with excessive 
radio frequency interference (RFI), which left 76 candidates to search.  
Since it is likely that a pulsar's signal would be highly accelerated, these 
beams were processed separately with a very fine acceleration search.  Appendix~A 
shows a list of all 76 HMXBs that were searched.

\subsection{An SMC Point Source}

We conducted a deep targeted radio search of SNR J0047.2$-$7308 which
was centred on the X-ray point source identified within the SNR (see
\citet{dwc+01}). These observations were conducted to look for radio
pulsations from a putative pulsar at that location. Two separate
search observations of duration 5.3 and 5.6 hours were conducted with
the Parkes 64-m radio telescope in February 2008. These observations used 
the centre beam of the multibeam receiver with a centre frequency of 1390~MHz 
and 0.5~MHz filters giving a total bandwidth of 256~MHz \citep{mlc+01}.  The 
observing setup and the observations themselves were identical to the search 
observation reported in detail by \citet{cld+09} which targeted
XTE~J0103$-$728, with the exception of the integration times and the
sampling time used. In the radio observations reported here, a
sampling time of 80 $\mu$s was employed to preserve sensitivity to
millisecond pulsations from a young pulsar.

\subsection{High Resolution LMC Survey}

New surveys of the Large and Small Magellanic Clouds have recently been 
initiated using a new high resolution backend, the Berkeley-Parkes-Swinburne
data recorder (BPSR), located at the Parkes Telescope \citep{kjv+10}.
These surveys were performed using the multibeam receiver at Parkes with a
centre frequency of 1374~MHz and a bandwidth of 340~MHz.  Each
pointing was observed for 8600~s and the data were sampled every
64~$\mu$s (see Ridley $\&$ Lorimer, in preparation).  These surveys of
the LMC and SMC began in May 2009 and we plan to continue them until 2016.  
They are $\sim20\%$ complete with 49 out of 209 pointings observed and we 
report on the results to date.

\section{DATA REDUCTION}\label{sec:reduction}

The Parkes Multibeam data were searched multiple times with different 
search parameters. We first used the \textsc{sigproc} processing 
package\footnote{http://sigproc.sourceforge.net}
to search 60 DMs from $0-300$~pc~cm$^{-3}$ using a minimum 
signal-to-noise ratio (S/N) of 9.0, and this search was sensitive to pulsars 
with periods from 10~ms to 10~s.  A time-domain acceleration search, as 
implemented in the \textsc{sigproc} program {\tt seek}, was
performed over the range $\pm$50~m/s$^2$ using intervals of 1~m/s$^2$.
Most pulsars in binary systems have accelerations up to 5~m/s$^2$ \citep{clf+00}, 
however, some, such as PSR B1744$-$24A with $a$=33~m/s$^2$ \citep{ljmsd90}, 
can be significantly greater.  To minimize the drifting of the signal across 
multiple Fourier bins due to the trial acceleration differing from the true acceleration, 
we chose a step size so that $\Delta a = cP/T_{\rm{obs}}^2$.  With the 
8400~s observations of the Parkes Multibeam Survey, the acceleration search 
was fully sensitive to all pulsars in binary systems having spin periods greater 
than 235~ms.  The data were also searched for single bursts of radio emission having 
a pulse width between 64~$\mu$s and 65~ms.  A minimum flux density threshold for a 
single pulse search can be calculated \citep[see][]{cm03} as
\begin{equation}
\label{eq:single}
S_{\rm{min}}=\frac{S_{\rm{sys}}~(\rm{S}/\rm{N})_{\rm{min}}}{W}\sqrt{\frac{W}{n_p~\Delta f}},
\end{equation}
where $W$ is pulse width, $S_{\rm{sys}}$ is the system equivalent flux density (34~Jy averaged 
across all beams of the multibeam receiver), $n_{\rm{p}}$ is the number of polarisations summed 
(2 for our case), and $\Delta f$ is the receiver bandwidth.  This made the search sensitive to 
giant bursts (of width 65~ms) having fluxes greater than 54~mJy.

A deeper search was also performed using the \textsc{sigproc} processing package and searched  
160 DMs over a range of $0-800$~pc~cm$^{-3}$.  The Fourier S/N threshold used 
was 5.0, and all candidates above that were folded. However, after using the 
\textsc{presto}\footnote{http://www.cv.nrao.edu/~sransom/presto} program {\tt prepfold}, only 
candidates with a resulting {\tt prepfold} sigma 8.0 and greater were inspected. This allowed 
us to detect a candidate that might appear weak in the FFT but was then enhanced in the fold.  
Since sigproc conducts an incoherent sum of the harmonics in the Fourier search, and the 
folded data are coherently summed, a pulsar candidate can improve its signal-to-noise ratio 
significantly in the folded data and reveal an otherwise invisible pulsar. This method is, 
however, more computationally-demanding.  An acceleration search was also performed with a range 
between $\pm$10~m/s$^2$ using intervals of 0.1~m/s$^2$ which was fully 
sensitive to all pulsars with spin periods greater than 23.5~ms.  

Survey beams containing the 76 HMXBs were 
searched using periodicity, single pulse and acceleration searches.  DMs up to 
800~pc~cm$^{-3}$ (160 different DMs) were searched with candidates having a S$/$N 
greater than 5.0 being considered.  Accelerations between $\pm$20~m/s$^2$ with 
step sizes of 0.01~m/s$^2$ were tried, making this search fully sensitive to all binary 
pulsars having spin periods greater than 2.35~ms.

The pointed search observations of SNR~J0047.2$-$7308 were analyzed to look for
a radio periodicity from a possible pulsar. After excision of RFI via the 
\textsc{presto} program {\tt rfifind}, each of the two data sets were dedispersed 
using a range of DMs from 0 to 800~pc~cm$^{-3}$, which easily encompassed the expected DM 
range for pulsars in the SMC \citep{mfl+06}. Each separate dedispersed time 
series was Fourier transformed, and the resulting spectra were filtered and
harmonically summed, then searched for candidate
periodicities. Candidate signals were then checked by dedispersing and
folding the data at DMs and periods near the candidate values. Each
data set was also searched for dispersed impulsive signatures in case
the source was an intermittent pulsar or rotating radio transient
source \citep{mll+06}. The analysis closely followed the Fourier and
single pulse search analysis of similar data described by
\citet{cld+09}. No promising candidates were found in the Fourier or
single pulse search of either data set.

\section{RESULTS}\label{sec:results}

\begin{figure*}
\psfig{file=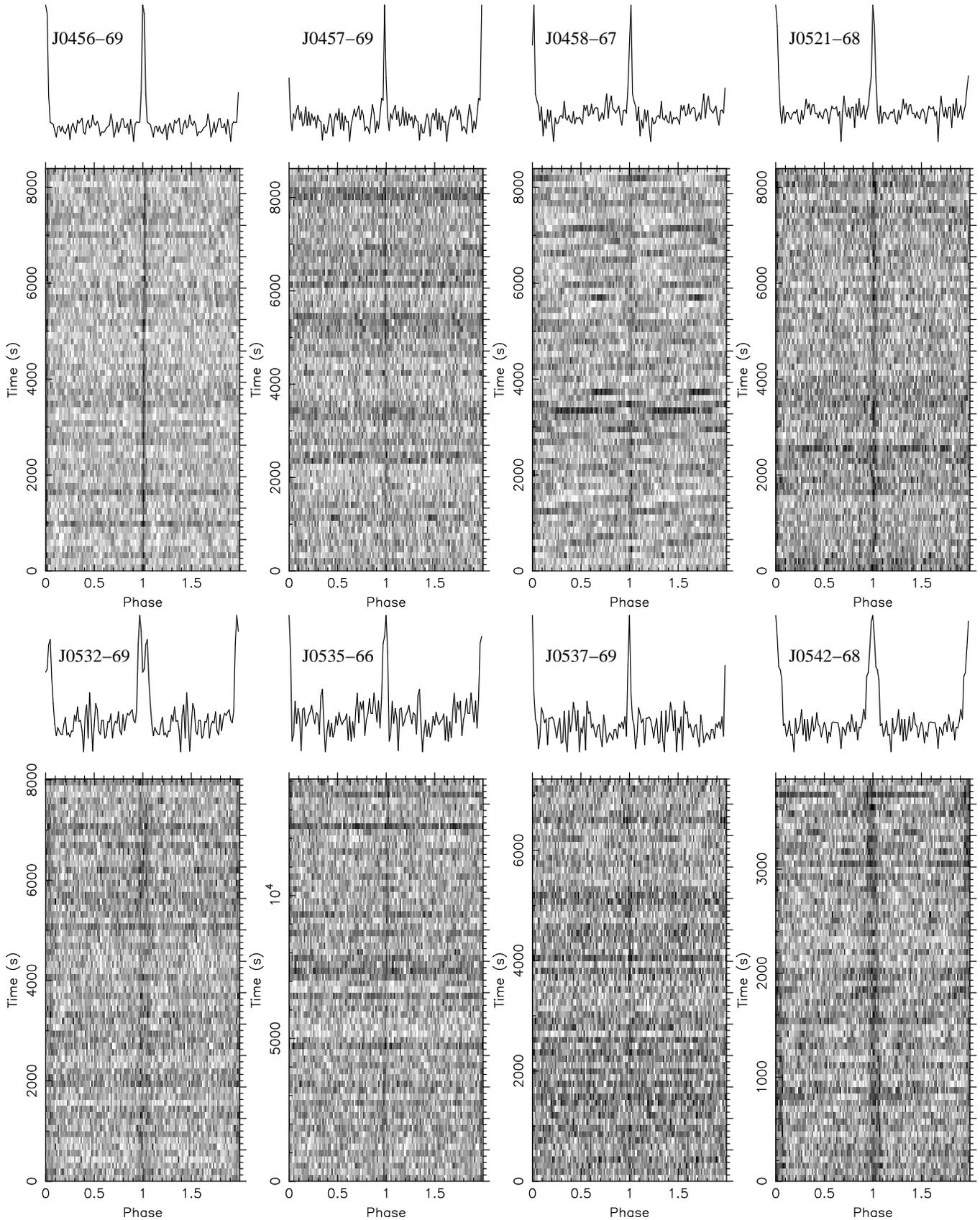,angle=0,width=18cm}
\caption{Discovery observations and integrated pulse profiles of the 
eight new LMC pulsars. For each pulsar, the grey scale shows pulse intensity
as a function of pulse phase and observation time. The sum of these
data result in the integrated profile shown at the top of each plot. Two pulse
phases are shown in each case for clarity.}
\label{fig:profs}
\end{figure*}

Four pulsars were discovered in the reprocessing of the Parkes 
Multibeam Survey of the Magellanic Clouds data, while one pulsar 
was found in the search of HMXBs.  An additional three pulsars 
were found in the initial processing of the high resolution LMC 
survey data.  These pulsars were confirmed in separate 
observations at the Parkes Radio Telescope between 14 Jan 2011 
and 24 June 2012.  Some general properties of these eight new 
radio pulsars can be found in Table~\ref{tb:newpsr}, while their 
time-resolved and integrated pulse profiles are shown in 
Figure~\ref{fig:profs}.  

\subsection{Properties of 8 New Pulsars}

These 8 new LMC pulsars encompass a wide range of periods and 
dispersion measures.  The periods range from 1.133~s for J0458$-$67 
down to 112~ms for J0537$-$69.  The 112~ms pulsar is the second shortest
period known for a radio pulsar in the LMC after PSR B0540$-$69.  It is also possible that 
it could be a young pulsar. Further timing observations are now
planned for this and the other new discoveries.

The range of dispersion measures of these pulsars is significant.  Three 
of the newly discovered have DMs less than 100~pc~cm$^{-3}$.  However, 
it is likely that they are all associated with the LMC since their DMs 
are all $>$70~pc~cm$^{-3}$, higher than the four lowest DM pulsars in the LMC, and 
well above the expected Galactic contribution to the DM in this direction ($\sim$25~pc~cm$^{-3}$; \citet{cl02}).  
PSR J0537$-$69, on the other hand, has a DM of 273~pc~cm$^{-3}$.  This is by far the highest 
DM of any known Magellanic Cloud pulsar (almost twice as high as the highest previously
known dispersion measure of 146~pc~cm$^{-3}$ for PSR B0540), and indicates that it is likely the 
most distant LMC pulsar yet discovered.  

\begin{table*}
\begin{center}
\caption{List of newly discovered LMC pulsars, including the barycentred 
period, DM, epoch of discovery, integration time, and discovery survey. PM refers to the Parkes Multibeam 
Survey of the Magellanic Clouds reprocessing, BPSR is the High Resolution Survey, and HMXB is the 
targeted search of HMXBs.  Parentheses represent the uncertainty in the last digit quoted.  Right 
ascension is given in units of hours, minutes, and seconds, and the declination is given in units 
of degrees and arcminutes.  The uncertainty in both RA and DEC is $\pm7$ arcminutes.  The uncertainty 
in DM was determined using {\tt prepfold} plots, which is part of the \textsc{presto} package.}
\label{tb:newpsr}
\begin{tabular}{lcccccccc}
\hline
           & RA          & DEC      & $P$           & DM                     & EPOCH & T$_{\rm{int}}$ & Survey     & S/N\\
PSR   & (J2000)  & (J2000) & (s)             & (pc cm$^{-3}$) & (MJD)    &  (s)                    &                   &        \\
\hline
J0456$-$69 & 04 56 30 & $-$69 10  & 0.117073051(15) & 103(1) & 52038 & 8300 & PM & 29.2\\
J0457$-$69 & 04 57 02 & $-$69 46  & 0.231390388(73) & 91(1)  & 55383 & 8600 & BPSR & 11.3\\
J0458$-$67 & 04 58 59 & $-$67 43  & 1.1339000(18)   & 97(2)  & 51810 & 8300 & PM & 10.8\\
J0521$-$68 & 05 21 44 & $-$68 35  & 0.43342070(30)  & 136(4) & 51871 & 8300 & PM & 19.3\\
J0532$-$69 & 05 32 04 & $-$69 46  & 0.57459786(70)  & 124(1) & 55420 & 7920 & BPSR & 9.6\\
J0535$-$66 & 05 35 40 & $-$66 52  & 0.210524357(30) & 75(1)  & 51393 & 14400 & HMXB & 7.1\\
J0537$-$69 & 05 37 43 & $-$69 21  & 0.112613212(20) & 273(1) & 55420 & 7200 & BPSR & 8.5\\
J0542$-$68 & 05 42 35 & $-$68 16  & 0.42518900(72)  & 114(5)  & 51975 & 3960 & PM & 18.6\\
\hline
\end{tabular}
\end{center}
\end{table*}

\subsection{SNR~J0047.2$-$7308}

No radio pulses were observed from the point-like source
in the supernova remnant J0047.2$-$7308.  The flux density limits on
radio pulsations from the radio search depend on the period and duty
cycle of any pulsations emitted by the source. For values that are
typical for young pulsars (a 5\% duty cycle and a spin period that is
between a few tens and a few hundreds of milliseconds), the 1400 MHz
flux density upper limit to pulsed emission from the blind Fourier
search is $\sim$~30~$\mu$Jy.  For a distance of 60 kpc to the SMC
\citep{HHH05} and assuming a 1~sr beaming fraction for any radio
emission, this corresponds to 1400 MHz radio luminosity of $\sim$~100
mJy kpc$^{2}$. The limits on single-pulse emission are comparable to
those reported for the radio search of XTE~J0103$-$728 in the SMC
\citep{cld+09}.

\section{DISCUSSION}\label{sec:discussion}

The known LMC pulsar population has now increased to 23 pulsars 
with the addition of these 8 newly discovered pulsars.  This is
a $50\%$ increase in the total number of LMC pulsars.  In this 
section we discuss the DMs and flux densities of our new 
pulsars as compared to the previously known pulsars.  
Table~\ref{tb:allpsr} shows the properties of all currently known LMC pulsars.
\begin{center}
\begin{table*}
\caption{Complete list of all 23 known rotation powered LMC pulsars.}
\label{tb:allpsr}
\begin{tabular}{lcccc}
\hline
               & P   & DM  & S$_{1400}$ & Discovery \\
PSR Name       & (ms)& (pc cm$^{-3}$) & (mJy) & Paper \\
\hline
J0449$-$7031   & 479 & 65  &  0.14 & \cite{mfl+06} \\
J0451$-$67     & 245 & 45  &  0.05 & \cite{mfl+06} \\
J0455$-$6951   & 320 & 94  &  0.25 & \cite{mmh+91} \\
J0456$-$69     & 117 & 103 &  $>$0.15 & This work \\
J0456$-$7031   & 800 & 100 &  0.05 & \cite{mfl+06} \\
J0457$-$69     & 231 & 91  &  $>$0.05 & This work \\
J0458$-$67     & 1133& 97  &  $>$0.07 & This work \\
J0502$-$6617   & 691 & 68  &  0.25 & \cite{mmh+91}\\
J0519$-$6932   & 263 & 119 &  0.32 & \cite{mfl+06} \\
J0521$-$68     & 433 & 136 &  $>$0.12 & This work \\
J0522$-$6847   & 674 & 126 &  0.19 & \cite{mfl+06} \\
J0529$-$6652   & 975 & 103 &  0.30 & \cite{mhah83} \\
J0532$-$6639   & 642 & 69  &  0.08 & \cite{mfl+06} \\
J0532$-$69     & 574 & 124 &  $>$0.05 & This work \\
J0534$-$6703   & 1817& 94  &  0.08 & \cite{mfl+06} \\
J0535$-$66     & 210 & 75  &  $>$0.03 & This work \\
J0535$-$6935   & 200 & 89  &  0.05 & \cite{ckm+01} \\
J0537$-$69     & 112 & 273 &  $>$0.04 & This work \\
J0537$-$6910   & 16  & $-$ &  $-$  & \cite{mgz+98} \\
J0540$-$6919   & 50  & 146 &  0.02 & \cite{ssh84}  \\
J0542$-$68     & 425 & 97  &  $>$0.14 & This work \\
J0543$-$6851   & 708 & 131 &  0.22 & \cite{mfl+06} \\
J0555$-$7056   & 827 & 73  &  0.21 & \cite{mfl+06} \\
\hline
\end{tabular}
\end{table*}
\end{center}

\subsection{Flux Densities}

\begin{figure}
\psfig{file=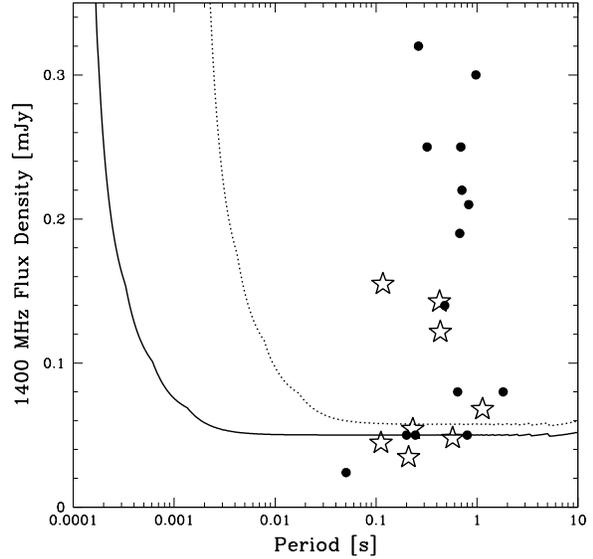,width=8cm}
\caption[Flux Density versus Pulse Period]{The 1400~MHz flux 
densities of all LMC pulsars are shown as a function of 
their pulse periods.  The 14 previously known radio pulsars (PSR J0537$-$6910 is 
not included) are represented by filled circles while the 8 new pulsars and the 
lower limits of their fluxes are denoted by stars.  The dashed line corresponds 
to the minimum flux threshold of the \cite{mfl+06} survey and the solid line is 
the minimum flux threshold of the new high resolution SMC and LMC surveys as 
calculated following the procedure described in \citet{mlc+01}.  The increased 
sensitivities of the new survey are due to the increase in bandwidth, decrease 
in sampling time, and increase in number of frequency channels.  The two pulsars 
below the survey thresholds (PSRs B0540$-$69 and J0535$-$66) were discovered 
using longer integration times.\\}
\label{fig:sens}
\end{figure}

A comparison of the flux densities of the previously discovered 
(from \citet{mfl+06}) and newly discovered pulsars is found in Figure~\ref{fig:sens}.  
We note that our improvements in computational power and search 
algorithms have enabled us to detect some pulsars with flux 
densities higher than the previous survey's minimum flux.  The flux 
values for the eight new pulsars were obtained via the radiometer equation,
\begin{equation}
\label{eq:rad}
S=\frac{T_{\rm{sys}}~(\rm{S}/\rm{N})}{G~\sqrt{n_{\rm{p}}~t_{\rm{int}}~\Delta f}}~\sqrt{\frac{W}{P-W}},
\end{equation}
where $G$ is the telescope gain, $n_{\rm{p}}$ is the number of polarisations 
summed, $T_{\rm{sys}}$ is the system temperature of the telescope, and $W$ and $P$ are 
the pulse width and period.  For these calculations we use $G$=0.735~K/Jy for the centre beam, 
$G$=0.690~K/Jy for the inner ring, and $G$=0.581~K/Jy for the outer ring (see \cite{mlc+01}), $n_{\rm{p}}$=2, 
and $T_{\rm{sys}}$=24~K.  The pulse width, $W$, is assumed to be 0.05~$P$.  Since the pulsars' positions 
are only known to within the beam radius, these quoted values are actually 
lower limits on their flux measurements.

\subsection{Dispersion Measures}

\begin{figure}
\psfig{file=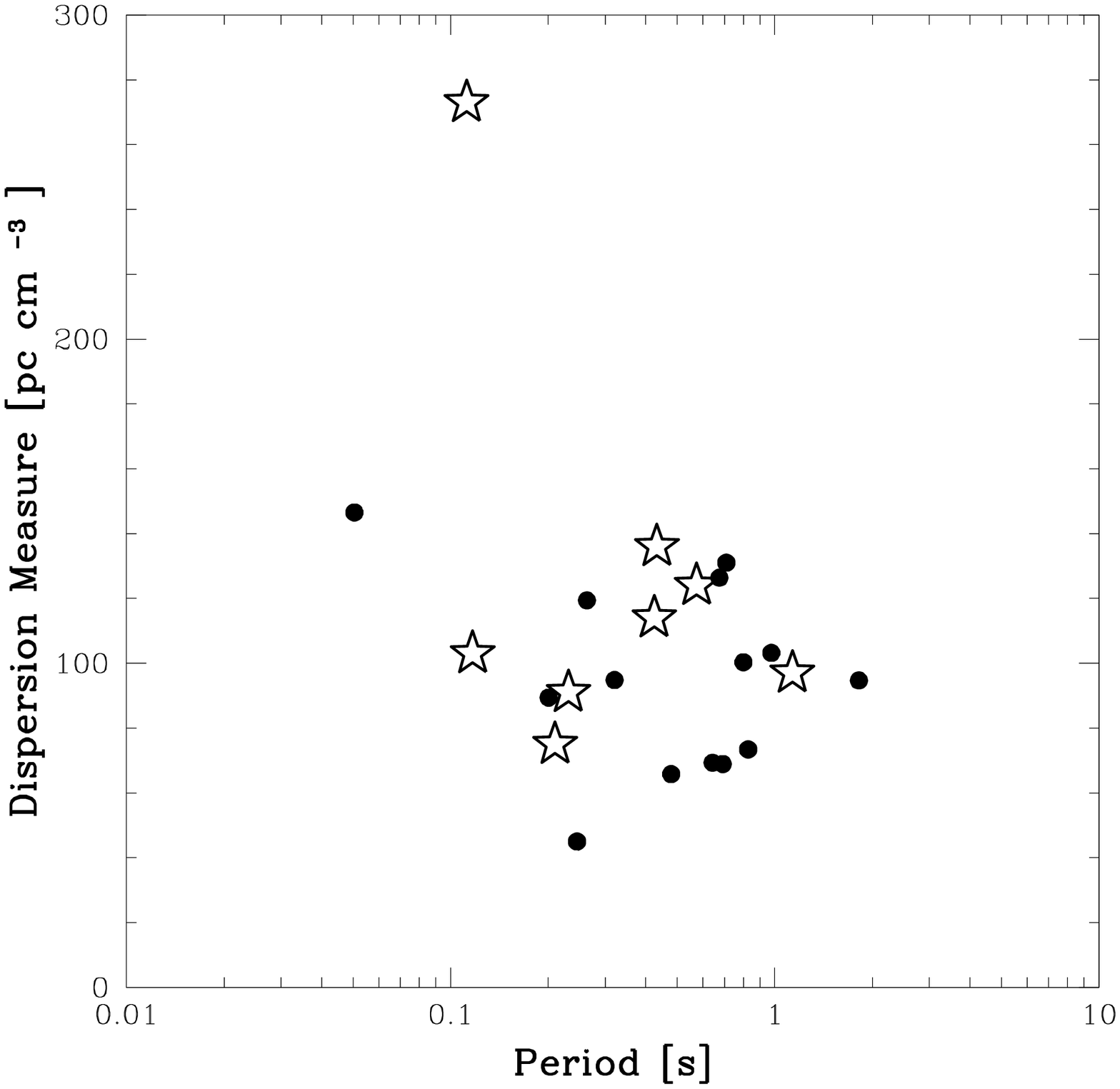,width=8cm}
\caption[Dispersion Measure versus Pulse Period]{The DMs of all LMC 
pulsars (with the exception of J0537$-$6910, which has not been radio 
detected) are displayed as a function of spin period.  Filled circles 
represent previously known pulsars and stars correspond to the new 
discoveries.\\}
\label{fig:dm}
\end{figure}

Most of the pulsars in both the Large and Small Magellanic 
Clouds have dispersion measures within the range of $65-150$~pc~cm$^{-3}$ \citep{mfl+06}.
PSR~J0537$-69$ was discovered to have a DM of 273~pc~cm$^{-3}$ which 
makes it the most highly dispersed pulsar located in the direction of 
the LMC and probably the most distant.  The DMs of all LMC pulsars 
(with the exception of the X-ray pulsar, J0537$-$6910) are plotted in Figure~\ref{fig:dm}.

\subsection{The population of LMC pulsars}

Using the current sample of pulsars in the LMC
we applied the Bayesian techniques developed by \citet{clmb13} to
constrain the number of potentially detectable LMC pulsars.  Here, we
take prior knowledge of the flux densities of the LMC pulsars and the
mean and standard deviation ($\mu$ and $\sigma$) of the log-normal luminosity
function \citep{fk06}, the use of which is justified below in Section \ref{sec:ldist},
to determine the population size ($N$). We follow
the procedure and used the code detailed in \citet{clmb13} to carry
out this analysis.  Since the diffuse radio flux of the LMC is not uniquely 
connected to the pulsar population, our analysis did not make use of the 
techniques described in Section 2.3 of Chennamangalam et al., where the 
diffuse flux of the millisecond pulsars is used to constrain globular cluster 
pulsar populations.  The X-ray pulsar, J0537$-$6910, does not have a 
measured radio flux and hence is not included in our analysis.  
Likewise, two pulsars (J0535$-$6935 and B0540$-$6919) were detected 
in targeted searches using much longer integration times than the 
surveys discussed here.  They were also excluded yielding a total of 
20 pulsars for this analysis.

\begin{figure}
\psfig{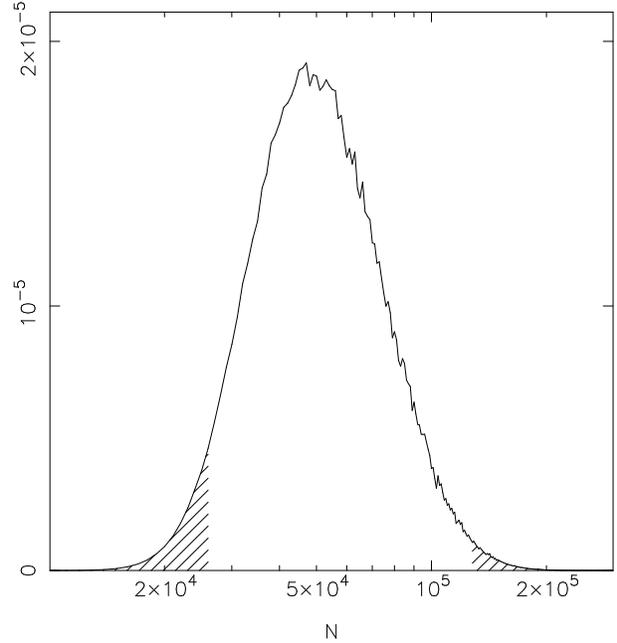}
\caption[Posterior on $N$ for Narrow Priors]{The posterior probability 
density function for the total number of potentially observable pulsars.  
The hatched areas represent regions outside of the 95\% credible interval.}
\label{fig:narrow}
\end{figure}

We assumed a distance of $49.97\pm1.3$~kpc to the LMC \citep{pgg+13} 
and a minimum survey sensitivity of 0.05 mJy.  Following 
\citet{blt+11}, we used the results of \citet{rl10a} to constrain the 
luminosity function priors to be flat in the ranges 
$-1.19 < \mu < -1.04$ and $0.91 < \sigma < 0.98$.  This results in a 
posterior probability density function (see Figure~\ref{fig:narrow}) for 
the total number of potentially observable pulsars with a median of 
57000$^{+70000}_{-30000}$ where the upper and lower limits are given by 
a 95\% credible interval.  

In an effort to further explore the luminosity function and 
underlying pulsar population we also tried a wide range of priors 
for $\mu$ and $\sigma$.  In this analysis we allowed $\mu$ to vary 
between $-$2.0 and 0.5 and $\sigma$ to vary between 0.2 and 1.4 according 
to \citet{blc11}.  The resulting constraints on the number of pulsars 
and the luminosity function are summarized in Table~\ref{tb:bayes}.  The 
constraints on the total number of pulsars for these wide priors are 
consistent with the narrow prior results, albeit with large uncertainties.  
The currently observed population of LMC pulsars is not large enough to 
uniquely constrain the shape of the pulsar luminosity function.  More 
sensitive surveys with future instruments would undoubtedly help to sample 
the fainter regions of the luminosity function and result in tighter constraints.
\begin{center}
\begin{table}
\caption{Median values for $N$, $\mu$, and $\sigma$ for our Bayesian analysis
of the LMC pulsar population and luminosity function.  For the narrow priors 
case the quoted values for $\mu$ and $\sigma$ only represent the range of priors 
that were used in the analysis.}
\label{tb:bayes}
\begin{tabular}{lccc}
\hline
       & N & $\mu$ & $\sigma$ \\
\hline
WIDE   & 21000$^{+71000}_{-20000}$ & -1.55$^{+1.55}_{-0.45}$ & 1.14$^{+0.22}_{-0.28}$ \\
\\
NARROW & 57000$^{+70000}_{-30000}$ & -1.12$^{+0.07}_{-0.07}$ & 0.95$^{+0.03}_{-0.04}$ \\
\end{tabular}
\end{table}
\end{center}

\subsection{Luminosity Distribution}
\label{sec:ldist}

\begin{figure*}
\psfig{file=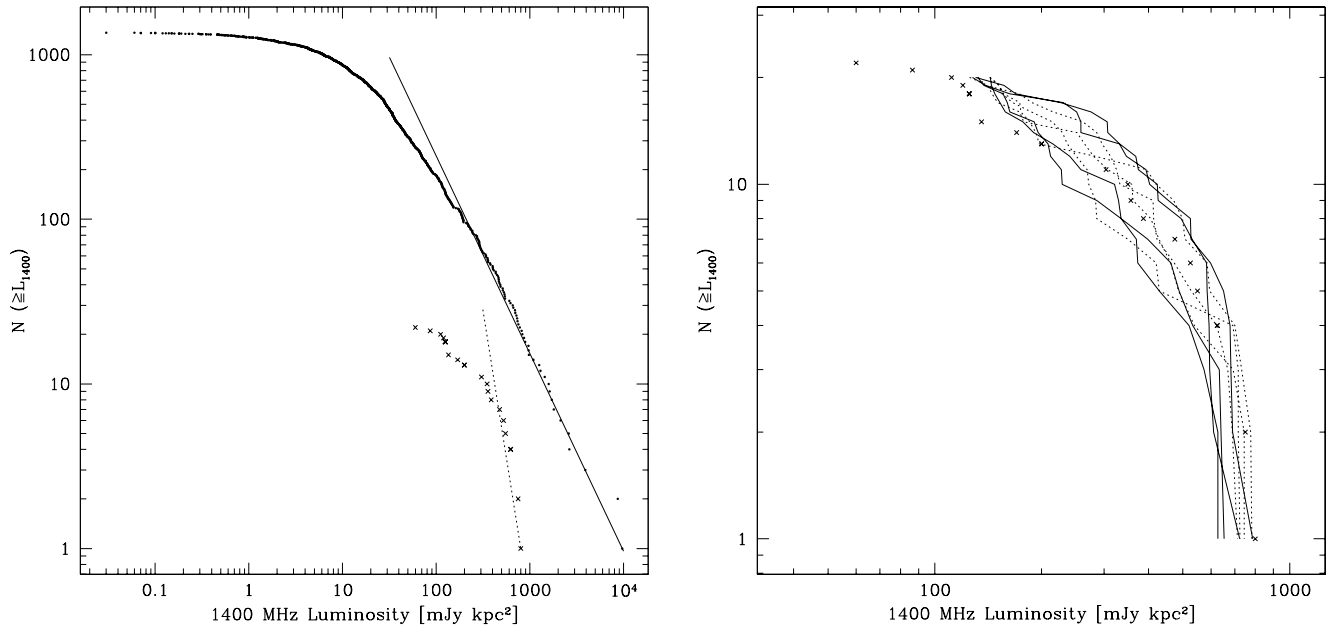,angle=0,width=18cm}
\caption[Luminosity Distributions]{The cumulative distribution function 
for both Galactic and LMC pulsars.  The left plot shows the Galactic pulsars (dots) 
and power law fit of $\alpha=-1.2$ and the LMC pulsars (represented by an X) with the 
corresponding power law fit of $\alpha=-3.6$.  The right plot shows only the LMC 
pulsars with 4 simulated populations using the power law distribution (solid lines) and 
4 simulated populations using the log-normal distribution (dotted lines).}
\label{fig:lum}
\end{figure*}

The above analysis assumes the log-normal luminosity distribution found from studies
of Galactic pulsars \citep{fk06}.
An interesting question to address with the growing sample of pulsars in the LMC is whether
their luminosity distribution is indeed consistent with the Galactic population. To quantify this,
we approximate the high-luminosity tail of the distribution as a power law, where the number of pulsars
($N$) as a function of luminosity ($L$) can be written as follows:
\begin{equation}
\textrm{log}~N=\alpha~\textrm{log}~L + C.
\end{equation}
Here $C$ is a constant and $\alpha$ represents the slope of the distribution \citep{lfl+06}.
We obtain a slope of $\alpha=-1.2$ for Galactic pulsars with $L>30$~mJy~kpc$^2$, and 
a slope of $\alpha=-3.6$ for LMC pulsars with $L>125$~mJy~kpc$^2$ (see Figure~\ref{fig:lum}).
At first glance, then, it appears that the slope and maximum luminosity of the two samples
are markedly different.

To investigate this apparent discrepancy between the Galactic and LMC luminosity 
distributions we simulated the luminosities of pulsars found in the LMC and compared them 
with the observed distribution.  We first generated 20 pulsars using a power law model 
with $\alpha=-1.2$\footnote{Similar results are found if we adopt a flatter slope
of $\alpha=-0.7$ as found by \citet{lfl+06}.} and plotted the resulting cumulative distribution function (CDF) along 
with the actual observed luminosities. We then simulated another set of 20 pulsars 
using the log-normal luminosity function values of $\mu=-1.1$ and $\sigma=0.9$ \citep{fk06}.  We 
ran 4 trials of each simulation type and plotted the results in Figure~\ref{fig:lum}.
As can be seen, the luminosities of the simulated pulsars do not vary significantly from the 
observed luminosities.  We conclude that, given the present statistics, the luminosity function
of normal pulsars in the LMC is consistent with that of their counterparts in the Galaxy.  Our
assumption of the log-normal luminosity function in the previous section therefore appears
to be completely justified by the present sample. We attribute the steeper slope of the observed LMC 
pulsars as a bias due to the smaller sample size. The higher maximum luminosity seen for
Galactic pulsars may be due in part to significant distance uncertainties associated with 
the Cordes \& Lazio electron density model which are not present in the LMC sample due to
our adoption of a common distance from \citet{pgg+13}. A cap on the luminosity
distribution, based on the LMC sample, seems to be approximately 1000~mJy~kpc$^2$.

\subsection{Small Magellanic Cloud}

Our reprocessing also searched the areas of the sky populated by the 
Small Magellanic Cloud.  However, no new pulsars were detected.  The smaller 
size of the cloud inherently leads to a fewer number of stars being present, 
hence the likelihood of detecting a pulsar is diminished.  \citet{rl10b} 
estimated that there are 40\% fewer pulsars in the SMC than in the LMC, 
or $\rm{N}_{\rm{SMC}}=0.6~\rm{N}_{\rm{LMC}}$.  Further decreasing our odds 
of a new discovery is the increased distance to the SMC of 60~kpc \citep{HHH05}.  
Since $S=L\times d^2$, the observed flux of an SMC pulsar as compared to an 
LMC pulsar is given by $S_{\rm{SMC}}=S_{\rm{LMC}}\big(\frac{48.3}{60}\big)^2$, meaning that 
the observed flux is reduced by 35\%.  The five currently known SMC pulsars are 
the most distant pulsars confirmed to date.

\section{CONCLUSIONS}\label{sec:conclusions}

With the 8 new pulsars presented in this work there are now  
23 known spin-powered pulsars known in the LMC, representing a 50\% 
increase in the population.  Based on these results, using a Bayesian
analysis, we constrain the total population of potentially observable
pulsars in the LMC to be 57000$^{+70000}_{-30000}$  (95\% credible
interval). The luminosity function of LMC pulsars is consistent with that
of normal pulsars in the Galaxy. Our results suggest that the maximum
1400-MHz radio luminosity is 1000~mJy~kpc$^2$.
Further work is being done to complete the 
high resolution BPSR survey of the LMC with the goal of establishing a 
deep and comprehensive survey of the entire galaxy. We hope to 
further increase the total number of known LMC pulsars and find the 
first extragalactic millisecond pulsars (see, e.g., \citet{rl10b}).

We are also conducting a BPSR survey of the SMC (Ridley \& Lorimer, in 
preparation) to find fainter and more rapidly rotating pulsars in 
that galaxy.  After observing 12 out of 73 planned pointings, no new 
pulsars have been detected, but a similar goal exists of performing 
an exhaustive search for pulsars in the SMC.

\section*{ACKNOWLEDGMENTS}

The Parkes radio telescope is part of the Australia Telescope which is
funded by the Commonwealth of Australia for operation as a National
Facility managed by CSIRO.  We thank Andrew Jameson, Matthew Bailes and Mike
Keith for assistance with the BPSR data acquisition system.
This work made use of the facilities of the ATNF Pulsar Catalogue. Summer 
support for SRB and computer resources at WVU used during this project were 
made possible by a WV EPSCoR Challenge Grant. DRL acknowledges support from 
the Research Corporation for Scientific Advancement as a Cottrell Scholar 
and current support from Oxford Astrophysics while on sabbatical leave.
Student work at F\&M was supported by the Hackman scholarship fund.  We also 
thank the referee, Matthew Bailes, for helpful comments. 

\bibliographystyle{mn2e}
\bibliography{references}

\appendix
\section{High-mass X-ray Binaries searched for radio pulsations in the LMC and SMC}

\begin{table*}
\caption{Catalog of 76 HMXBs in the LMC and SMC searched for radio pulsations 
and dispersed radio bursts.  The X-ray type refers to whether the 
source is a pulsar (P), transient X-ray source (T), or ultra-soft X-ray spectrum (U)
(see \citet{lph05}).  The RA is given in units of hours, minutes, and seconds, and 
the DEC is given in units of degrees, arcminutes, and arcseconds.}
\label{tb:hmxb}
\begin{center}
\begin{tiny}
\begin{tabular}{llllllll}
\hline \hline
Name                   & RA (J2000)  & DEC (J2000) & Orbital Period (d) & X-ray Type & Spect Type  & Spin Period (s) & Alt Name              \\
\hline
RX J0032.9-7348        & 00 32 56.10 & -73 48 19.0 &         &          & Be                     &              &                             \\
RX J0041.2-7306        & 00 41 16.40 & -73 06 41.0 &         &          &                        &              &                             \\
AX J0042.0-7344        & 00 42 04.80 & -73 44 58.0 &         &          &                        &              &                             \\
RX J0045.6-7313        & 00 45 37.90 & -73 13 54.0 &         &          & Be                     &              &                             \\
RX J0047.3-7312        & 00 47 23.42 & -73 12 27.3 &  48.800 &   P      & B2e                    &     263.6400 & AX J0047.3-7312             \\
AX J0048.2-7309        & 00 48 14.90 & -73 10 03.0 &         &          & Be                     &              &                             \\
RX J0048.5-7302        & 00 48 34.50 & -73 02 30.0 &         &          & Be                     &              & XMMU J004834.5-730230       \\
RX J0049.2-7311        & 00 49 13.84 & -73 11 36.7 &         &   P      & Be                     &       9.1321 & XMMU J004913.8-731136       \\
RX J0049.5-7310        & 00 49 29.92 & -73 10 58.0 &  91.500 &          & Be                     &              & XMMU J004929.9-731058       \\
RX J0049.7-7323        & 00 49 42.00 & -73 23 15.0 & 394.000 &   P      & B1-3Ve                 &     755.5000 & XMMU J004942.3-732313       \\
XTE J0050-732\#1        & 00 50 00.00 & -73 16 00.0 & 189.000 &   P     &                     &      16.6000 &                            \\
XTE J0050-732\#2        & 00 50 00.00 & -73 16 00.0 &         &   P     &                       &      51.0000 &                            \\
RX J0050.7-7316        & 00 50 44.70 & -73 16 05.0 &   1.416 &   P      & B0III-Ve               &     323.0000 & AX J0051-733                \\
RX J0050.9-7310        & 00 50 57.60 & -73 10 07.9 &         &          & Be                     &              & AX J0050.8-7310             \\
XTE SMC46              & 00 51 00.00 & -73 18 00.0 &         &   TP     &                        &      46.4000 &                             \\
RX J0051.3-7250        & 00 51 19.60 & -72 50 44.0 &         &          & Be                     &              &                             \\
AX J0051.4-7227        & 00 51 25.40 & -72 27 29.0 &         &          &                        &              &                             \\
AX J0051.6-7302        & 00 51 39.90 & -73 02 58.0 &         &          &                        &              &                             \\
XTE J0051-727          & 00 51 42.00 & -72 45 00.0 &         &   TP     &                        &     293.9000 &                             \\
RX J0051.9-7255        & 00 51 54.20 & -72 55 36.0 &         &          & Be                     &              &                             \\
XTE J0052-725          & 00 52 09.10 & -72 38 03.0 &         &   TP     &                        &      82.4000 &                             \\
2E0051.1-7304          & 00 52 52.40 & -72 48 30.0 &         &          & B0e                    &              &                             \\
RX J0052.9-7158        & 00 52 59.20 & -71 57 58.0 & 200.000 &   TPU    & Be                   &     169.3000 & 2E 0051.1-7214              \\
CXOU J005323.8-722715  & 00 53 23.80 & -72 27 15.0 & 125.000 &   P      & Be                     &     138.0000 & RX J0053.4-7227             \\
XTE SMC95              & 00 53 24.00 & -72 49 18.0 & 280.000 &   TP     &                        &      95.0000 &                             \\
XTE Position A         & 00 53 54.00 & -72 26 42.0 &         &   P      &                        &      89.0000 &                             \\
1WGA J0053.8-7226      & 00 53 55.00 & -72 26 47.0 & 139.000 &   TP     & B1-B2III-Ve            &      46.6300 & XTE J0053-724               \\
RX J0054.5-7228        & 00 54 33.20 & -72 28 09.0 &         &          & Be                     &              &                             \\
H 0053-739             & 00 54 36.20 & -73 40 35.0 &         &   TP     & B1.5Ve                 &       2.3700 & SMC X-2                     \\
AX J0054.8-7244        & 00 54 55.88 & -72 45 10.5 & 261.000 &   TP     & O9Ve                   &     503.5000 & RX J0054.9-7245             \\
XTE J0055-724          & 00 54 56.17 & -72 26 27.6 & 123.000 &   TP     & B0-B1III-V             &      58.9690 & 1SAX J0054.9-7226           \\
XMMU J005517.9-723853  & 00 55 18.44 & -72 38 51.8 &         &   P      & O9V                    &     701.6000 & RX J0055.2-7238             \\
XTE J0055-727          & 00 55 24.00 & -72 42 00.0 &  34.800 &   TP     &                        &      18.3700 &                             \\
CXOU J005527.9-721058  & 00 55 27.70 & -72 10 59.0 &         &   P      &                        &      34.0800 & RX J0055.4-7210             \\
XMMU J005605.2-722200  & 00 56 05.24 & -72 22 00.9 &         &   P      & Be                     &     140.1000 & 2E0054.4-7237               \\
XMMU J005615.2-723754  & 00 56 15.20 & -72 37 54.0 &         &          & Be                     &              &                             \\
XMMU J005724.0-722357  & 00 57 24.00 & -72 23 57.0 &         &          &                        &              &                             \\
AX J0057.4-7325        & 00 57 26.80 & -73 25 02.0 &         &   P      &                        &     101.4500 & RX J0057.4-7325             \\
CXOU J005736.2-721934  & 00 57 36.20 & -72 19 34.0 &  95.300 &   P      & Be                     &     564.8300 & XMMU J005735.7-721932       \\
RX J0057.8-7202        & 00 57 48.40 & -72 02 42.0 &         &   P      & Be                     &     280.4000 & AX J0058-7203               \\
CXOU J005750.3-720756  & 00 57 50.30 & -72 07 56.0 &         &   P      &                        &     152.1000 & RX J0057.8-7207             \\
RX J0058.3-7216        & 00 58 20.70 & -72 16 18.0 &         &          &                        &              & AX J0058.3-7217             \\
RX J0059.2-7138        & 00 59 11.30 & -71 38 45.0 &         &   TUP    & B1IIIe                 &       2.7632 &                             \\
1XMMU J005921.0-722317 & 00 59 21.04 & -72 23 16.7 &         &   P      & B0                     &     202.0000 & RX J0059.3-7223             \\
XMMU J010030.2-722035  & 01 00 30.23 & -72 20 35.1 &         &          & Be                     &              &                             \\
RX J0101.6-7204        & 01 01 37.56 & -72 04 18.7 &         &          & Be                     &              &                             \\
XTE J0103-728          & 01 03 24.00 & -72 43 00.0 &         &   TP     &                        &       6.8482 &                             \\
RX J0103.6-7201        & 01 03 37.57 & -72 01 33.2 &         &   P      & O5Ve                   &    1323.2000 &                             \\
RX J0104.5-7221        & 01 04 35.60 & -72 21 43.0 &         &   T      & Be                     &              &                             \\
RX J0105.7-7226        & 01 05 41.60 & -72 26 17.0 &         &          &                        &              & XMMU J010541.5-722617       \\
AX J0113.0-7246        & 01 13 05.50 & -72 46 25.0 &         &          &                        &              &                             \\
RX J0117.6-7330        & 01 17 41.40 & -73 30 49.0 &         &   TP     & B0.5IIIe               &      22.0700 &                             \\
AX J0127.8-7307        & 01 27 48.00 & -73 07 39.0 &         &          &                        &              &                             \\
RX J0456.9-6824        & 04 56 54.10 & -68 24 35.0 &         &          &                        &              &                             \\
RX J0457.2-6612        & 04 57 12.40 & -66 12 10.0 &         &          &                        &              &                             \\
RX J0501.6-7034        & 05 01 23.90 & -70 33 33.0 &         &          & B0Ve                   &              & CAL 9                       \\
RX J0502.9-6626        & 05 02 51.60 & -66 26 25.0 &         &   TP     & B0Ve                   &       4.0635 & CAL E                       \\
RX J0512.6-6717        & 05 12 41.80 & -67 17 23.0 &         &          &                        &              &                             \\
RX J0524.2-6620        & 05 24 12.70 & -66 20 50.0 &         &          &                        &              &                             \\
RX J0527.3-6552        & 05 27 23.70 & -65 52 35.0 &         &          &                        &              &                             \\
RX J0529.4-6952        & 05 29 25.90 & -69 52 11.0 &         &          &                        &              &                             \\
RX J0529.8-6556        & 05 29 48.40 & -65 56 51.0 &         &   TP     & B0.5Ve                 &      69.5000 &                             \\
RX J0531.2-6607        & 05 31 13.80 & -66 07 03.0 &  25.400 &   TP     & B0.7Ve                 &      13.7000 & EXO 053109-6609.2           \\
RX J0531.5-6518        & 05 31 36.10 & -65 18 16.0 &         &          & B2V                    &              &                             \\
RX J0532.4-6535        & 05 32 25.30 & -65 35 09.0 &         &          &                        &              &                             \\
2A 0532-664            & 05 32 46.10 & -66 22 03.0 &   1.400 &   P      & O8 III                 &      13.5000 & LMC X-4                     \\
RX J0535.0-6700        & 05 35 05.90 & -67 00 16.0 & 241.000 &   T      & B0Ve                  &              &                             \\
1A 0535-668            & 05 35 41.20 & -66 51 52.0 &  16.700 &   PT     & B0.5 IIIe              &       0.0690 & 1A 0538-66                  \\
RX J0535.6-6651        & 05 35 41.60 & -66 51 58.0 &         &          &                        &              &                             \\
RX J0535.8-6530        & 05 35 53.80 & -65 30 34.0 &         &          &                        &              &                             \\
1H 0538-641            & 05 38 56.30 & -64 05 03.0 &   1.700 &   U      & B2.5 Ve                &              & LMC X-3                     \\
XMMU J054134.7-682550  & 05 41 34.70 & -68 25 50.0 &         &          &                        &              &                             \\
RX J0541.5-6833        & 05 41 37.10 & -68 32 32.0 &         &          & B0III                  &              & RX J0541.6-6832             \\
1SAX J0544.1-7100      & 05 44 06.30 & -71 00 50.0 & 286.000 &   TP     & B0Ve                 &      96.0800 & RX J0544.1-7100             \\
H 0544-665             & 05 44 15.60 & -66 34 59.0 &         &          & B0 Ve                  &              &                             \\
RX J0546.8-6851        & 05 46 48.30 & -68 51 47.0 &         &          &                        &              &                             \\
\hline
\end{tabular}
\end{tiny}
\end{center}
\end{table*}

\end{document}